\documentclass[journal]{IEEEtran}
\usepackage{latexsym,amssymb,exscale,graphicx,amsfonts,amsmath}
\usepackage[utf8]{inputenc}
\usepackage{multirow,multicol}
\usepackage{epstopdf}
\usepackage{epsfig} 
\usepackage{setspace}
\usepackage[table]{xcolor}
\usepackage{xcolor,colortbl}
\usepackage{lipsum}
\usepackage{microtype}
\usepackage{stfloats}
\usepackage{cite}
\usepackage{enumitem}
\begin{document}
\newcommand{\nrc}[1]{\textcolor{red}{NRC: #1}}
\newcommand{\gop}[1]{\textcolor{blue}{GOP: #1}}
\newcommand{\ting}[1]{\textcolor{magenta}{TH: #1}}
	
	\title{Topology Estimation Following Islanding and its Impact on Preventive Control of Cascading Failure}
	
	\author{Sai Gopal Vennelaganti,~\IEEEmembership{Student Member,~IEEE,}
		Nilanjan Ray Chaudhuri,~\IEEEmembership{Senior Member,~IEEE,}
		Ting He,~\IEEEmembership{Senior Member,~IEEE,}
		and Thomas La Porta,~\IEEEmembership{Fellow,~IEEE}
		\thanks{Financial support from NSF Grant Award ECCS 1836827 is gratefully acknowledged.}}
	\maketitle

	\begin{abstract}
	Knowledge of power grid's topology during cascading failure is an essential element of centralized blackout prevention control, given that multiple islands are typically formed, as a cascade progresses. Moreover, academic research on interdependency between cyber and physical layers of the grid indicate that power failure during a cascade may lead to outages in communication networks, which progressively reduce the observable areas. These challenge the current literature on line outage detection, which assumes that the grid remains as a single connected component. We propose a new approach to eliminate that assumption. Following an islanding event, first the buses forming that connected components are identified and then further line outages within the individual islands are detected. In addition to the power system measurements, observable breaker statuses are integrated as constraints in our topology identification algorithm. The impact of error propagation on the estimation process as reliance on previous estimates keeps growing during cascade is also studied. Finally, the estimated admittance matrix is used in preventive control of cascading failure, creating a closed-loop system. The impact of such an interlinked estimation and control on that total load served is studied for the first time. Simulations in IEEE-118 bus system and 2,383-bus Polish network demonstrate the effectiveness of our approach.
	\end{abstract}

	
	\IEEEpeerreviewmaketitle
	\vspace{-10pt}
	\section{Introduction} \label{sec:intro}
	\IEEEPARstart{C}{ascading} failure in a power system can lead to disruption of many services and cause loss of power 
 	comparable to natural disasters \cite{disaster}. Although a low probably event, its associated high socio-economic impact merits research for a better understanding of this phenomena and preventive control strategies. In this context, reference  \cite{prevent} presented studies on interdependency between cyber and  physical  layers  of  the  grid  indicating  that  large scale cascading failures in power network may lead to outages in communication networks of supervisory control and data acquisition (SCADA), which disrupts the monitoring and control of the power system. The same paper also proposed a preventive control based on optimal generation reduction and load shedding under progressively reducing controllability and observability due to communication failures, but assumed accurate knowledge of the system topology (i.e. breaker statuses) nonetheless.  
 	The main focus of our paper is to fill this gap by estimating topology as the number of observable breaker statuses keeps diminishing with time, and study the impacts of estimation on preventive control of cascading failure.
	
	Despite research interest in line outage 
	estimation in general~\cite{overbye, overbye2,dependence,sparseidenti,external,entropy, entropypmu,badadata,compressive,quickest,SVMforesti},  
	not many papers have dealt with the specific challenges that come with cascading failure in power grids modeled as a cyber-physical system. We note that almost all of the papers in~\cite{overbye, overbye2,dependence,sparseidenti,external,entropy, entropypmu,badadata,compressive,quickest,SVMforesti} assume a DC quasi steady state (DC-QSS) model for the purpose of problem formulation, as is assumed in our work as well. Although this DC-QSS model is used for cascading failure studies \cite{prevent,DC2,DC3,DC4,DC5}, when validating the proposed approach, a more realistic AC-QSS model \cite{benchmarking} has been used in this paper. 
	
	Three aspects that are unique to line outage detection during cascading failure are: (i) complete system information 
	(including breaker statuses and line flows) is available before the outage, but only partially known as the cascade progresses, (ii) multiple line outages may occur in succession, and (iii) formation of many islands within the power grid due to these line outages. The first challenge has been adequately addressed in the current literature, while much less attention has been paid to the latter two. Tate et-al, \cite{overbye, overbye2} has formulated the problem of line outage detection (PLOD) in the external power system. Here, the complete system information is known before outage via the system data exchange and only limited data from phasor measurement units (PMUs) of local network is available after outage. Although the exact problem setup of this paper is different, the methods developed in \cite{overbye} can still be applied here. Similarly, the techniques proposed in this paper can also be used for PLOD in external grid via PMU data. The only difference is that the proposed method uses line flows, whereas \cite{overbye} uses phase angles. Since the line flows and phase angles are linearly related in DC model, depending on the requirement, PLOD formulation can be easily switched, see Section \ref{sec:line detection} for further details. 
	
	Following the work of \cite{overbye}, many have proposed improvements or better algorithms and techniques to solve the same problem. For example, \cite{dependence} took a graphical approach by modeling the PLOD through a Gaussian Markov random field, while, \cite{sparseidenti} took advantage of sparsity. The PLOD was represented as an integer programming problem in \cite{external}. References \cite{entropy, entropypmu} employed cross-entropy based optimization technique to solve the same. Further, \cite{badadata} proposed methods to deal with bad data in PMU measurements, and \cite{compressive} introduced a new sparse formulation to better solve the PLOD. In \cite{quickest}, the statistical properties of the small random fluctuation has been exploited to detect line outages using a linearized model and \cite{SVMforesti} applied support vector machine classification on time series data from PMU to identify and locate the line outages. However, these approaches are limited to a one or two line outages.

	None of the literature discussed above has tested their techniques during a cascading failure, especially with multiple line outages. The techniques could have worked if not for a strong requirement of connectedness of the system. To the best of our knowledge, only \cite{dobson} has extended the PLOD proposed in \cite{overbye} to deal with islanding. However, that too was limited to a single line outage and no specific optimization was proposed to reach the solution. This leads to the third and final challenge, which is unanswered in the literature, `how to detect line outages in the event of multiple island formations?' This paper answers the question by formulating an optimization problem, which upon solving identifies the buses within each island. Then a modified version of PLOD proposed in \cite{overbye} 
	is formulated as least absolute shrinkage and selection operator (LASSO) problem \cite{lasso}, to detect any further line outages within the individual islands. Moreover, observable breaker statuses are also integrated as constraints into this process to improve the accuracy of topology estimation. 
	
	Instead of verifying the proposed topology estimation just for different sets of initial line outages, the proposed estimation is verified at every step of the cascading failure. This enables us to study the method in presence of error propagation, i.e., as reliance on previous estimates keeps growing. Finally, the estimated admittance matrix is used in preventive control of cascading failure formulated in \cite{prevent}, creating a closed-loop system. The impact of such an interlinked estimation and control on final load served is also studied in IEEE-118 bus system 
	and 2,383-bus Polish system \cite{matpow}. 
	
		\vspace{-10 pt}
	\section{Modeling of Power System} \label{sec:model}
	%
	%
	
	In literature, power system cascading failure has been mostly studied through QSS models \cite{prevent,DC2,DC3,DC4,DC5,benchmarking}, since dynamic model \cite{dyn2} is expensive for studying the impact of various initiating outages. The simplest of all models is the DC QSS model \cite{prevent,DC2,DC3,DC4,DC5}, 
	which enables the usage of well-understood mathematical tools to solve the complex problem of topology estimation when islanding occurs. 
	Although, to have a practical usage, there is definitely a need to expand the developed technique to the AC-QSS model \cite{benchmarking}  
	or to the Decoupled-DC-QSS model; for this complex problem, the DC-QSS model serves as a reasonable stepping stone towards that final goal. Thus motivated, in this paper, all the proposed methodologies are developed using the DC-QSS model and are validated using the AC-QSS model. Extension to more complex models is left to future work.
	
		\vspace{-10pt}
	
	\subsection{Physical Layer Modeling of the Power System}
	Consider a power system modeled by an undirected graph ${\cal{G}}_{p} := ({\cal N},\cal{E})$, where ${\cal N}$ is the set of buses (nodes) connected by transmission lines (edges) denoted by a set ${\cal E}:= \{u,v\} \subseteq {\cal N} \times {\cal N}$. The phase angle of voltage of bus $u$ is $\theta_u$ and corresponding power injection is $p_{u}$. The reactance of line $l = (u,v)$ is $x_{uv}=x_{vu}$ and the power flowing from bus $u$ to bus $v$ is given by,  
	\begin{equation} \label{eq:lineflowsingle}
	\begin{aligned}
	\ell_{l} = \ell_{uv} = \frac{1}{x_{uv}} (\theta_u - \theta_v) 
	\end{aligned}
	\end{equation}
	
The power injected at a particular bus $u$ should be equal to sum of powers flowing in the lines that emerge out of the bus $u$. This implies,

	\begin{equation} \label{eq:power_injsingle}
	\begin{aligned}
	p_u  = \sum_{v \in {\cal N}_u}\frac{1}{x_{uv}} (\theta_u - \theta_v) 
	\end{aligned}
	\end{equation}
	where, ${\cal N}_u$ is the set of buses that are neighbors of the bus $u$. Note that, we are assuming that all the tap changers have been locked to a ratio of unity.
The two equations (\ref{eq:lineflowsingle}) and (\ref{eq:power_injsingle}) can be expanded to all the buses and links in the network ${\cal{G}}_{p}$ and can be compactly represented as follows,
	\begin{equation} \label{eq:power_flow_raw}
	\begin{aligned}
	\mathbf{p} = \mathbf{ M D M}^\mathsf{T} \boldsymbol{\theta} \\
	\boldsymbol{\ell} = \mathbf{D M}^\mathsf{T} \boldsymbol{\theta} \hspace{6 pt} 
	\end{aligned}
	\end{equation}
	Here, $\mathbf{p} := (p_u)_{u\in {\cal N}}\in \mathcal{R^{|\cal N|}}$, $\boldsymbol{\theta} := (\theta_u)_{u\in {\cal N}}\in \mathcal{R^{|\cal N|}}$, $\boldsymbol{\ell} := (\ell_{l})_{l\in {\cal{E}}}\in \mathcal{R}^{|\cal{E}|}$. Also, $\mathbf{D} := diag(d_1,...,d_{|\cal{E}|}) \in \mathcal{R^{|\cal{E}| \times |\cal{E}|}}$ s.t. $d_l = 1/x_{uv}$.
	The matrix $\mathbf{M}$ is the incidence matrix of the graph ${\cal{G}}_{p}$, whose columns comprise of vectors $\{ \mathbf{m}_l \}_{l = 1} ^{|\cal{E}|}\in \mathcal{R^{|\cal N|}}$. The entries of the vector $ \mathbf{m}_l $ are all $0$, except $u$th and $v$th entries, which are $1$ and $-1$, respectively. 
	The admittance matrix of the power system $\mathbf{B}$ can be constructed as follows,
	\begin{equation} \label{eq:admittance_mat}
	\begin{aligned}
	\mathbf{B}  = \mathbf{ M D M}^\mathsf{T}
	\end{aligned}
	\end{equation}
	Traditionally, the power injections $\mathbf{p}_i$ will be given and we solve for the flows $\boldsymbol{\ell}$. In that process, as a first step the following equation is solved for $\boldsymbol{\theta}$,
	\begin{equation} \label{eq:power_flow}
	\begin{aligned}
	\mathbf{p} = \mathbf{B} \boldsymbol{\theta} 
	\end{aligned}
	\end{equation} 
	Then, from the phase angles $\boldsymbol{\theta}$, the flows are calculated using (\ref{eq:power_flow_raw}). For a connected power system, the admittance matrix $\mathbf{B}$ is of rank ${|\cal N|}-1$ and results in multiple solutions for $\boldsymbol{\theta}$. To avoid this, one of the buses is chosen as a reference bus for each island and its angle is set to zero or any fixed angle. 
	
	\vspace{-10 pt}
	\subsection{Cyber Layer Modeling of the Power System}
The cyber layer of the power system consists of two main networks, (i) SCADA, and (ii) wide-area monitoring, protection, and control (WAMPAC). The later uses PMUs as sensors, and is rapidly growing. However, at this stage, it has limited observability of the power system. The former typically employs redundant data measurements, which are used for topology and state estimation. Therefore, in this paper, only the SCADA network is considered as part of cyber layer modeling.  
	
	\begin{figure}[h]
			\vspace{-5pt}
		\hspace{-5pt}
		\includegraphics[scale = 0.5, trim= 0.4cm 0.3cm 0.4cm 0.5cm, clip=true]{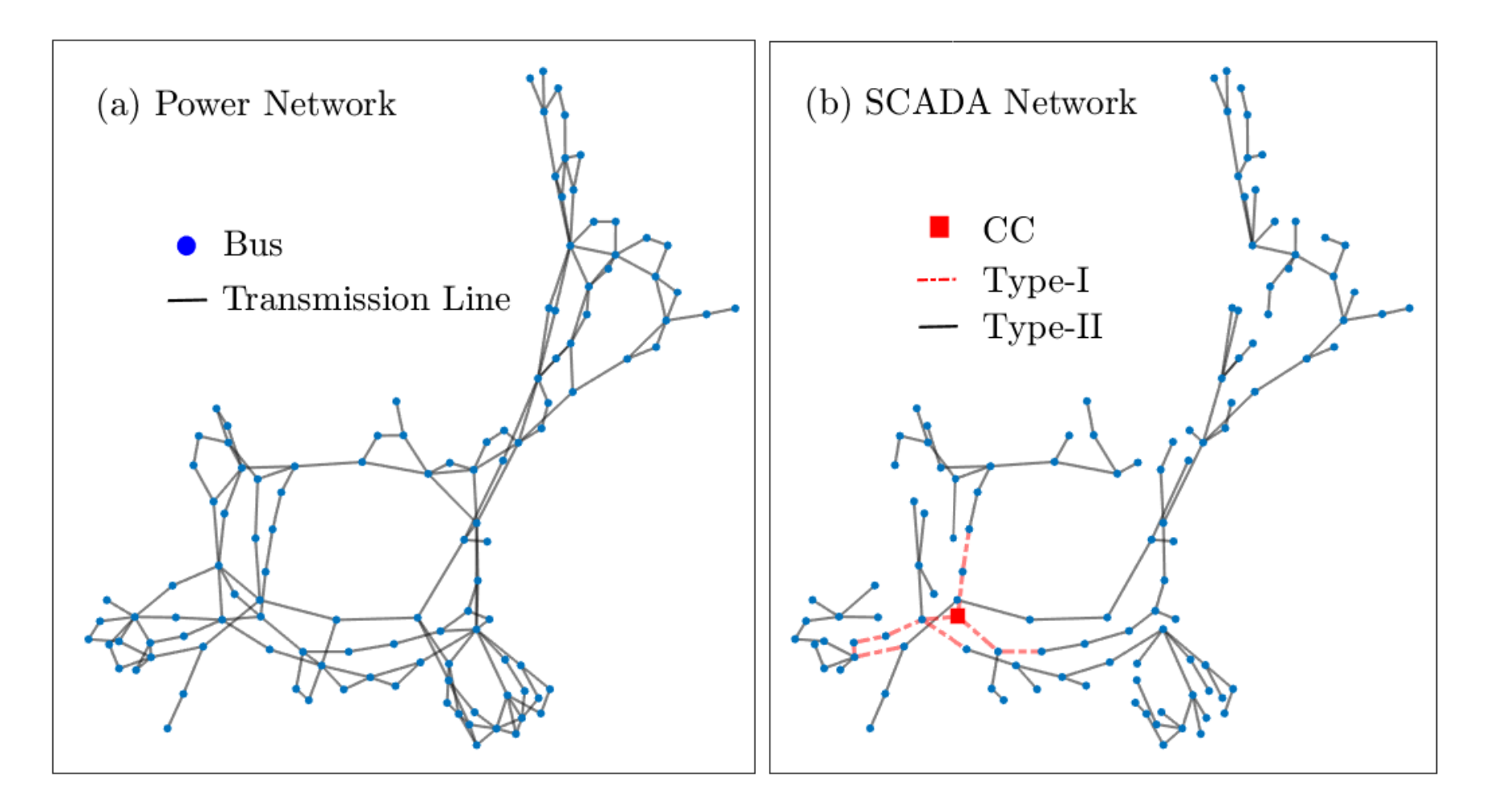}
		\vspace{-15pt}
		\caption{IEEE 118-bus system, (a) Physical-layer: power network graph and (b) Cyber layer: SCADA network graph.}
		\vspace{-5pt}
		\label{fig:con}
	\end{figure}  
	
To be conservative, the SCADA network topology is assumed to be a minimum spanning tree of the physical power network.
Each communication node is geographically collocated with a bus \cite{prevent}. As an example, power and communication graphs of IEEE 118-bus system are shown in Fig. \ref{fig:con} (a) and (b), respectively. 
All the sensors and actuators at the buses are connected to the control center (CC) node, which is highlighted in Fig. \ref{fig:con} (b) using a red square. We assume communication links of two types, Type-I link and Type-II link. The Type-I links (e.g. wireless communication links and fiber-optic channels with back up power in nodes) are robust and they will not go down if the corresponding transmission line trips. On the other hand, a Type-II link (e.g. power line carrier communication assuming backup is absent/disabled in the worst case) is susceptible to power line outages, i.e., it is assumed to go down if the corresponding power line trips.
For the purpose of case studies, it is assumed that only $10$, i.e., around $10\%$ of the links, are of Type-I in the IEEE 118-bus system, which is a conservative assumption. As highlighted in Fig. \ref{fig:con} (b), they are located around the CC. 
	
In this communication model, as long as a node is connected to the CC via some path, all the sensor information of the bus collected by that node is communicated to CC. This includes, power generated and/or consumed at the bus, voltage magnitude, line flows and breaker statuses of all the transmission lines incident to the bus. Note that unlike WAMPAC, the
phase angle information is not available here. Also, the generation and load consumption of buses connected to the CC through the communication network can be controlled.

\textit{Remark:} We made certain pessimistic assumptions regarding the SCADA network for conservative analysis of topology estimation and its implication on cascade prevention control under progressively reducing observability and controllability in extreme conditions. We remark that the proposed topology estimation approaches are in no way dependent on these assumptions.
\vspace{-10 pt}
	\subsection{Coupled Cascading Failure Model of Power System}
	\begin{figure*}[t]
		\vspace{-20 pt}
		\centering
		\includegraphics[trim = {2.1cm 0.8cm 0.5cm 0.8cm}, clip,width=0.86\textwidth]{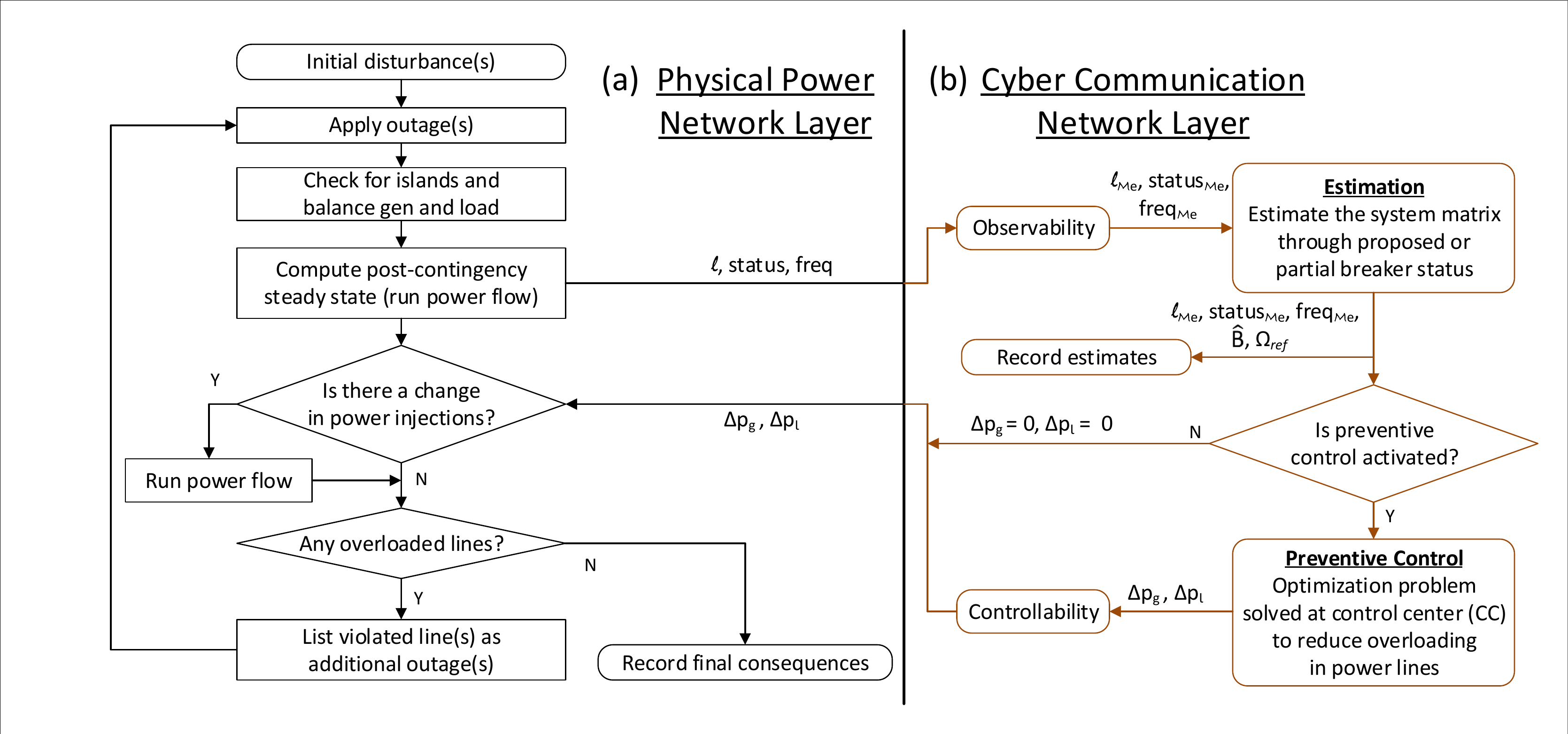}
		\vspace{-5 pt}
		\caption{Flow chart highlighting the coupled cascading failure model of power and communication networks.}
		\vspace{-15 pt}
		\label{fig:overall}
	\end{figure*}
	
Figure \ref{fig:overall} illustrates the cascading failure model of the physical power system and how it interacts with the cyber layer. Reference \cite{benchmarking} presents comparisons among various existing models of cascading failure of the physical power system. In this paper, a simple deterministic model is considered, where all the overloaded lines are simultaneously tripped. As shown in Fig.~\ref{fig:overall}, measured line flows ($\boldsymbol{\ell}_{{\cal M}e}$), frequencies (\textbf{freq}$\mathbf{_{{\cal M}e}}$), and breaker statuses (\textbf{status}$\mathbf{_{{\cal M}e}}$) with a progressive loss of observability are used to estimate the admittance matrix ($\mathbf{\hat{B}}$) through the proposed two-step approach described in Sections \ref{sec:island_detection} and \ref{sec:line detection}. Here, ${{\cal M}e}$ refers to measured quantities, and the term `observability' refers to the amount of information receivable by control center, given the state of the communication layer.
When preventive control is active, it takes $\mathbf{\hat{B}}$ as input for solving the optimization problem that generates set point changes to generation ($\Delta \mathbf{p}_g$) and load ($\Delta \mathbf{p}_l$) to contain the spread of cascade. Here, the term `controllability' refers to numbers of generation and load buses that can receive command to shed power, given the state of communication layer. In this \textit{closed-loop} scenario, the error in $\mathbf{\hat{B}}$ due to declining observability and reduced controllability stemming from communication failures jointly impact the effectiveness of the control action, which is studied in Section \ref{sec:Full_B}.
	
   \vspace{-10pt}
	\section{Identifying the Island Formations} \label{sec:island_detection}
	%
	%
	%

	The existing literature on topology estimation requires that the physical power system remains connected. However, in this paper there is no such requirement. In our proposed method, the topology estimation is treated as a two step process. As a first step, following any islanding, the buses that form the islands are identified. Next, within each island, the unobservable line outages are localized. This section describes the processes involved in the first step, i.e., island detection. 
	
	\begin{figure*}
		\vspace{-15pt}
				\hspace{-5pt}
		\includegraphics[scale = 0.33, trim= 0.5cm 0.6cm 0.4cm 0.5cm, clip=true]{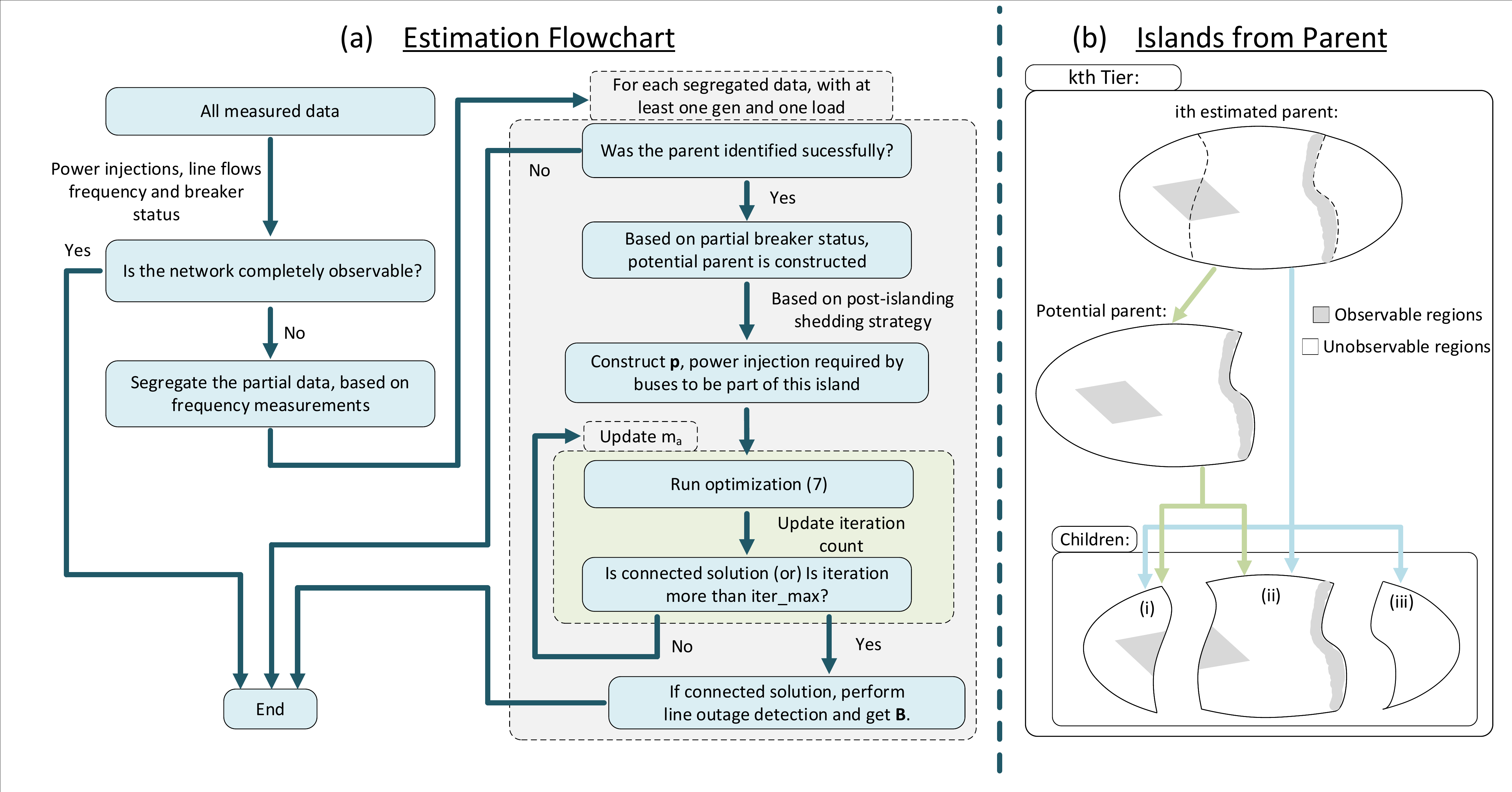}
		\vspace{-10pt}
		\caption{Overview of the entire estimation process: (a) Flowchart, (b) child islands formed from $i$th parent island of the $k$th tier, during the cascading failure.}
		\label{fig:flowchart}
		\vspace{-15pt}
	\end{figure*}  
	
	\vspace{-10 pt}
	\subsection{Prerequisites for Island Detection}
	
	The following assumptions are made to implement this method, which are listed below,
	\begin{enumerate}
		\item Complete information of the system, i.e., power injection at all buses and topology before failure is known.
		\item Future load consumption in unobservable parts can be quite accurately predicted with high probability. 
		\item Frequency measurement is precise enough to group observable buses into various islands. 
		\item No unknown outages of generators or loads take place due to dynamic instability or oscillatory behavior. 
		\item No unknown outages or trips occur from under voltage load shedding.
	\end{enumerate}
	Since the SCADA-based energy management system at the control center collect remote terminal unit and topology processor data to perform state estimation. This ensures the availability of bus power injections and the topology before cascading. 
	Also, the second assumption is supported by accurate load forecasting models
	that are currently used.
	If the frequency measurements (in Hz) are precise enough, then the observable buses from two different islands can be distinguished from each other. This is because different islands will have different frequencies with high probability, since the generation-load imbalance is likely to be different in each island upon their formation.  The last two assumptions stem from the fact that DC-QSS model can only capture outages caused by line overloads, but not dynamic and voltage related phenomenon, which can be represented by detailed dynamic models. Note that these two assumptions are omnipresent in PLOD literature~\cite{overbye, overbye2,dependence,sparseidenti,external,entropy, entropypmu,badadata,compressive,quickest,SVMforesti} -- further studies including these factors are left to future work.  
	

The process of topology estimation is explained in Fig. \ref{fig:flowchart} (a). Following any outages, we check for complete observability of the system. In the event of partial observability where estimation is needed, the measured data is segregated based on frequency measurements. From the 
islands that were estimated in the previous tier, a parent island, if one exists, is identified. Next, from the partial breaker status, a potential parent is constructed. Figure \ref{fig:flowchart} (b) illustrates this process with an example. In the $k$th tier, say the $i$th parent, which was successfully estimated -- breaks down into child islands (i), (ii) and (iii). Here, the observable parts are marked in gray in Fig. \ref{fig:flowchart} (b). The split between (i) and (ii) is not completely observable. However, from the partial breaker status, the split between (ii) and (iii) can be observed. Therefore, during the identification of island (i) and island (ii), the potential parent as highlighted in Fig. \ref{fig:flowchart} (b) is constructed and used. 
	
The most important requirement for island detection is that, at least one generator and one non-zero load bus should be observable from the island. This enables us to estimate the shedding in other buses that belong to the island of interest. Without loss of generality, a simple proportional shedding strategy is assumed in this paper. This means, if the generation is higher than load in a newly formed island, then generation is reduced in all generator buses in proportion to their current absolute values until the total generation equals load and vice versa. 
Therefore, if we can measure the reduction in one generator, then we can extend this information and estimate the required power injections in all other the generator buses -- if they were to be part of the same island. This is also true for the observable non-zero load buses.

\textit{Remark:} We note that the above generation and load shedding strategy is in no way limiting to our proposed approach. For any generic case, all we need is the knowledge of the governor droop coefficients of generators and the underfrequency load shedding scheme in the utility.

\vspace{-10pt}	
\subsection{Problem Statement of Island Detection}	
\vspace{-3pt}

Going back to the previous example of identifying island (i) in Fig. \ref{fig:flowchart} (b), let us say that the observable part of island (i) contains a few generators and non-zero load buses. From them, the required shedding in power injection of all the buses of potential parent shown in Fig. \ref{fig:flowchart} (b) can be computed. This is possible due to estimation performed in the previous tier, which means we are relying on accurate prediction of the parent. Therefore, the estimated power injections in the previous tier 
are known. Note that for the first tier, the parent is the whole power system and it is assumed to be completely observable.
	
Now that we know the required power injection of buses to be part of the island of interest, for example island (i), the next step is to identify the largest set of buses that satisfy the following constraints, namely, 

\noindent\emph{(a)} The required power injections of the set of buses should sum to zero.

\noindent\emph{(b)} All these buses should be connected with each other through the transmission lines or transformers in potential parent island.
\noindent\emph{(c)}  Based on frequency data, buses that are from the island of interest should be included and buses that are from other islands must be excluded from this set.

Finding the buses that satisfy the above constraints can be translated into solving the following nonlinear integer programming problem,
\vspace{-6pt}
	\begin{subequations} 
	\begin{align}
	\min_{x_i \in \{0,1\}} ~ & -\mathbf{1}_m^\mathsf{T} \mathbf{x} \label{eq:obj}  \\
	\textrm{s.t.} ~~~~ & |  \mathbf{p}^\mathsf{T} \mathbf{x}| \le \epsilon; \label{eq:con1}    \\
	& \mathbf{|p|}^\mathsf{T} \mathbf{x} \ge \text{min}(\{|p(i)| : p(i) \ne 0 \}) \label{eq:con2}  \\
	& \mathbf{lb} \le \mathbf{x} \le \mathbf{ub} \label{eq:con3} \\
	& \mathbf{\tilde{A}} = \bigg( \sum_{i = 1}^{m} (\mathbf{e_i e^\mathsf{T}_i} ) \mathbf{ e^\mathsf{T}_i x} \bigg) \mathbf{A} \bigg( \sum_{i = 1}^{m} (\mathbf{e_i e^\mathsf{T}_i} ) \mathbf{ e^\mathsf{T}_i x} \bigg)^\mathsf{T} \label{eq:con4} \\
	& \mathbf{\tilde{L}} =   \bigg( \sum_{i = 1}^{m} (\mathbf{e_i e^\mathsf{T}_i} ) \mathbf{ e^\mathsf{T}_i (\tilde{A} 1_n)}  \bigg) -\mathbf{\tilde{A}} \label{eq:con5}  \\
	& 1+ \text{rank}(\mathbf{\tilde{L}} ) \ge \mathbf{1}_m^\mathsf{T} \mathbf{x} \label{eq:con6} 
	\end{align}  \label{eq:nonlinearintprog}
	\end{subequations}
	 Here, the symbol $|.|$ is a size-preserving element-wise absolute-value operator, $\mathbf{e_i}$ is the unit vector along the $i$th direction, and $\mathbf{1}_m$ is an $m\times1$ vector whose each element is $1$. The elements of vector $\mathbf{p}$ are the required power injections of buses in the potential parent. $\mathbf{x}$ is a vector, whose elements are binary (0 or 1). The indices of elements in $\mathbf{x}$ whose values are ones, correspond to the buses that are included in the set, i.e., they form the island. The objective function tries to maximize the number of buses in the set. The first constraint (\ref{eq:con1}) ensures the sum of power injections is as close to zero as possible, since $\epsilon$ is set to a very small value ($\approx1e-10$). The second constraint (\ref{eq:con2}) eliminates the trivial solution of choosing only zero injection load buses. Based on frequency data, the elements of $\mathbf{x}$ for buses we know are in the island are forced to one by setting the corresponding lower bound $\mathbf{lb}$ entries to ones. Similarly, the entries of upper bound $\mathbf{ub}$ corresponding to the buses we know are not in the island are set to zeros. 
	
	The last three constraints (\ref{eq:con4}-\ref{eq:con6}) ensure that the solution can form a connected component without the need of any additional buses. Here, $\mathbf{A}$ is the adjacency matrix of the power network graph of the potential parent. Similarly, $\mathbf{\tilde{A}}$ is the adjacency matrix of the subgraph formed from the potential parent based on the entries of $\mathbf{x}$ that are selected. The matrix $\mathbf{\tilde{L}}$ is Laplacian of that subgraph constructed from $\mathbf{\tilde{A}}$. If the rank of this matrix $\mathbf{\tilde{L}}$ is one less than total number of activated entries of $\mathbf{x}$, then the buses in the set can form a single connected component, which is the island of interest. 
	
	\vspace{-4 pt}
	\subsection{Simplifications to Island Detection Problem}

\begin{figure}
		\vspace{-5pt}
			\hspace{10 pt}
		\includegraphics[scale = 0.47, trim= 4.5cm 6cm 5cm 6cm, clip=true]{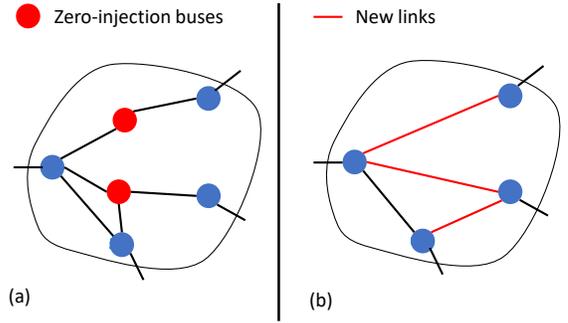}
		\vspace{-25pt}
		\caption{Elimination of zero-injection buses, (a) original power network graph and (b) reduced power network graph, with new links, highlighted in red.}
		\label{fig:zeroinj}
		\vspace{-18pt}
	\end{figure}

	Without fundamentally changing the problem, eliminating the zero-injection buses can simplify the problem. As shown in Fig. \ref{fig:zeroinj}, the zero-injection nodes highlighted in red can be replaced by links that connect the neighboring buses, which is again highlighted through a red link in Fig. \ref{fig:zeroinj} (b). This results in the formation of a reduced power network graph. The optimization problem then changes as follows, the matrix $\mathbf{A}$ is replaced by $\mathbf{\hat{A}}$, which is the adjacency matrix of the reduced power network graph of the potential parent. The vector $\mathbf{p}$ is replaced by $\mathbf{\hat{p}}$, which contains only the non-zero power injections in the reduced power network. The second constraint (\ref{eq:con2}) is also updated to $\mathbf{1}_n^\mathsf{T} \mathbf{x} \ge 1$, which eliminates the trivial solution of not choosing any bus (i.e., $\mathbf{x} = \mathbf{0}$).
	
The next simplifying modification is to ignore the connectedness condition in the modified optimization problem, which gives rise to a standard integer linear programming problem, 
	\begin{equation} \label{eq:theislanddet}
	\begin{aligned}
	\min_{x_i \in \{0,1\}}  ~~ & -\mathbf{1}_n^\mathsf{T} \mathbf{x}  \\
	\textrm{s.t.} ~~~~~  & |  \mathbf{\hat{p}}^\mathsf{T} \mathbf{x}| \le \epsilon ; ~~
	1 \le \mathbf{1}_n^\mathsf{T} \mathbf{x} \le m_a   \\
	& \mathbf{lb} \le \mathbf{x} \le \mathbf{ub}  \\
	\end{aligned}
	\end{equation}
	The problem can now be solved using several off-the-shelf solvers like, `Gurobi' \cite{gurobi}. This new problem is solved multiple times iteratively, until a desired solution is reached as illustrated in Fig. \ref{fig:flowchart}. Solving (\ref{eq:theislanddet}) with $m_a$ greater than length of $\mathbf{\hat{p}}$, gives us the biggest set, which definitely satisfies the constraint (a), mentioned earlier. We then verify if this solution satisfies the connectedness constraint (b). If it does not, the size of the solution will be restricted to one less than the previously obtained solution through appropriate choice of $m_a$. The optimization problem (\ref{eq:theislanddet}) is solved again for the next 
	iteration. The process is repeated until a connected solution is reached or the maximum number of solutions verified is hit. Once a final connected solution is found, all the neighboring zero-injection buses from the original potential parent are added to the list of nonzero injection buses that were identified.
	
		\vspace{-4pt}
	\subsection{Results} \label{sec:res:islandiden}
	
	\begin{table}[h]
	\vspace{-20pt}
		\caption{Percentage of accuracy of island detection, assuming the parents have been correctly estimated.}
		\vspace{-5pt}
		\hspace{-5pt}
		\begin{tabular}{cccc}
			\hline
			Test system                                  & \multicolumn{3}{c}{IEEE 118 bus}     \\ \hline
			\multicolumn{1}{c|}{\% out. (\# nodes)}       & 1\% (2) & 2\% (3)  & 3\% (4)     \\ \hline
			\multicolumn{1}{c|}{acc. / high obs.}    & 85.62/39.21 & 80.27/32.84 & 80.72/31.22 \\ \hline
			\multicolumn{1}{c|}{acc. / low obs.} &  69.29/11.37   &  67.63/12.17   & 63.83/11.32    \\ \hline
			\multicolumn{1}{c|}{tot. acc. / obs.} &  81.95/50.58   &  76.85/45.01   & 76.23/42.54    \\ \hline \hline
			Test system                                  & \multicolumn{3}{c}{Polish grid} \\ \hline
			\multicolumn{1}{c|}{\% out. (\# nodes)}       & 0.2\% (5) & 0.4\% (10)  & 0.6\% (15)    \\ \hline
			\multicolumn{1}{c|}{acc. / high obs.}    & 84.62/83.04 & 69.4/73.65 & 61.08/69.38 \\ \hline
			\multicolumn{1}{c|}{acc. / low obs.} &  -/-   &  100.00/0.24 & -/-\\ \hline
			\multicolumn{1}{c|}{tot. acc. / obs.} &  84.62/83.04 &  69.5/73.89   & 61.08/69.38  \\ \hline
			
		\end{tabular}  \label{tab:accuracyisland}
		\vspace{-7pt}
	\end{table}

	In this section, before discussing how line outages within the detected islands are identified, we will examine the effectiveness of the proposed island detection method in AC-QSS models. To that end, it is assumed that the parent island has been correctly estimated i.e., the parent island is same as the ground truth. This way, the reliance on previous estimates is ruled out, and the effectiveness of the estimation of child islands can be studied independently. Table \ref{tab:accuracyisland} lists the estimation accuracy in the IEEE $118$-bus system and the Polish grid test systems for various initial outage sizes and different observability levels. For each percentage of initial nodal outage, $100$ and $500$ random cases were considered in the $118$-bus and the Polish system, respectively. Also, `high obs.' indicates percentage of parent islands with at least $50\%$ observable nodes (including at least one generator bus and one nonzero load bus)  while `low obs.' represents the remaining fraction of parent islands with observability in at least one generator bus and one nonzero load bus. The remaining islands are unobserved and at least for the purpose of preventive control, it is not relevant to estimate the topology of those islands, because they are also uncontrollable. The first entry in the first row $85.62/39.21$ indicates that $39.21\%$ of newly formed islands have high observability
	and that $85.62\%$ of those $39.21\%$ islands have been correctly identified. As per intuition, higher observability has resulted in higher accuracy of island detection, which can be achieved through a resilient communication network. 

		\vspace{-0pt}
	\section{Line Outage Detection in  Identified Islands} \label{sec:line detection}
		\vspace{-0pt}
	%
	%
	
	\begin{figure}
		\vspace{-10pt}
			\hspace{5pt}
		\includegraphics[scale = 0.50, trim= 7cm 2cm 11.5cm 1cm, clip=true]{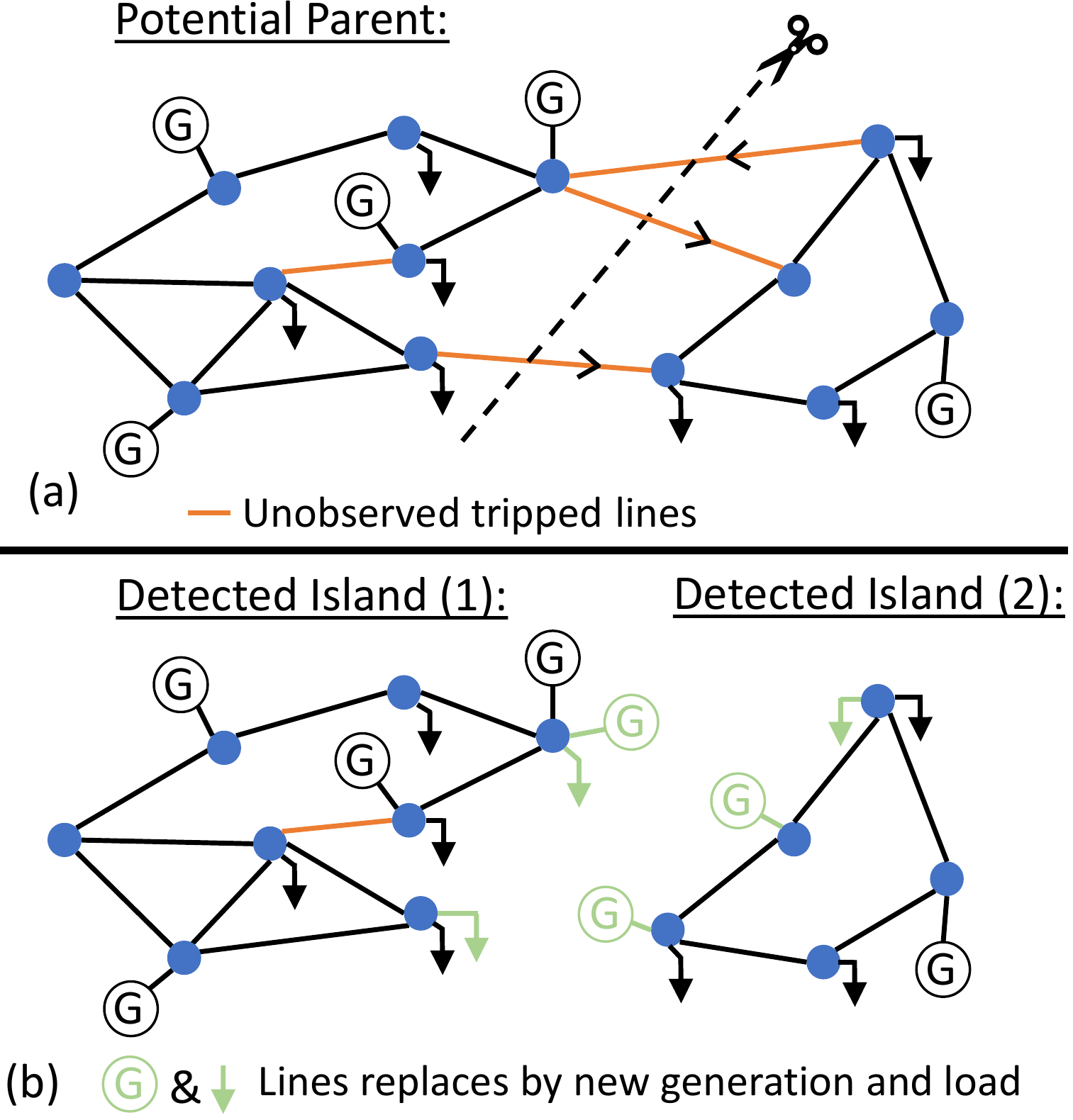}
		\vspace{-10pt}
		\caption{Replacing tripped lines between two islands with fictitious generation and load, (a) parent power system graph, and (b) graph of detected island assuming all unobservable lines within are still connected.}
		\label{fig:linedetect}
		\vspace{-18pt}
	\end{figure}
	
	Following the identification of the buses belonging to the island of interest, as illustrated in Fig. \ref{fig:flowchart}, we proceed to detect any line outages that occurred within the island. Although multiple papers have proposed line outage detection, in this paper we modified the existing approach to deal with bus power injection changes from gen-load balancing and boundary line outages. Also, we incorporated the available partial breaker statuses and used difference in line flows instead of phase angles for the estimation. Note that the line outage detection method trusts the island identification and assumes the identified island is accurate (be it accurate or not). Since the  power injection of all buses  were already estimated during island identification, the line flows and voltage phase angles of each island can be computed by solving load flow equations (\ref{eq:power_flow}).

	\vspace{-5pt}
	\subsection{Proposed Line Outage Detection Approach}	
	\vspace{-2pt}
	
	The first step before formulating and solving the problem of line outage detection is the construction of matrix $\mathbf{B}$ corresponding to the island of interest with the available information. As shown in Fig. \ref{fig:linedetect}, the parent will break down into smaller islands, which are detected through bus identification proposed in the previous section. 
	For each bus with incident line outages 
	at the boundary of a detected island and rest of the system, a generator and load duo is used to mimic the effect of the pre-outage line flows as shown in Fig. \ref{fig:linedetect} (b). Within the island, say Island (1), observable breaker statuses indicating tripped lines 
	are disconnected while the unobserved lines (some of which might be out) are all assumed to be connected. Now we construct an admittance matrix $\mathbf{B}$ for that island and assume that power injections are same as pre-outage injections, say $\mathbf{p}$. Since this is a newly constructed island, power flow equation (\ref{eq:power_flow}) can be solved to obtain phase angles $\boldsymbol{\theta}$ and line flows $\boldsymbol{\ell}$.
	
	Let $\mathbf{B}'$ be the actual (ground truth) admittance matrix of the island following the outage that needs to be estimated. 
	Assuming that island estimation is correct, the difference between $\mathbf{B}$ and $\mathbf{B}'$ is due to the unobserved line outages alone. Let us define the difference to be $\tilde{\mathbf{B}} := \mathbf{B} - \mathbf{B}'$. Let the post-outage power injections, phase angles and line flows be, $\mathbf{p}'$, $\boldsymbol{\theta}'$, and $\boldsymbol{\ell}'$, respectively. Similarly, let the difference between post- and pre-outage quantities be denoted by $\tilde{\mathbf{p}}$, $\tilde{\boldsymbol{\theta}}$, and $\tilde{\boldsymbol{\ell}}$. The difference 
	$\tilde{\mathbf{p}} := \mathbf{p}' - \mathbf{p}$ comes from shedding when the island is formed and 
	boundary lines are removed, which is mimicked by tripping of generation and/or load that were added to represent the boundary line flows, see Fig. \ref{fig:linedetect}. From (\ref{eq:power_flow}) we know that,
	\begin{equation} \label{eq:powinjdiff}
	\begin{aligned}
	\mathbf{B}\boldsymbol{\theta} + \tilde{\mathbf{p}}= \mathbf{B}'\boldsymbol{\theta}' = (\mathbf{B} -\tilde{\mathbf{B}} )\boldsymbol{\theta}'
	\end{aligned}
	\end{equation}
	Note that, except $\tilde{\mathbf{B}}$, all the other differences are of the form, `\textit{post-}' minus `\textit{pre-}'. Rearranging the above equation gives us,
	\begin{equation} \label{eq:powinjdiff2}
	\begin{aligned}
	\mathbf{B}\tilde{\boldsymbol{\theta}} = \tilde{\mathbf{B}} \boldsymbol{\theta}' + \tilde{\mathbf{p}} \hspace{19 pt}\\
	= \sum_{l \in \tilde{\cal E}} s_l \mathbf{m}_l + \tilde{\mathbf{p}} \\
	\end{aligned}
	\end{equation}
	here, $\tilde{\boldsymbol{\theta}} = \boldsymbol{\theta}' - \boldsymbol{\theta}$; $s_l := \mathbf{m}_l^\mathsf{T} \boldsymbol{\theta}'/x_l, \forall l \in \tilde{\cal E}$, and $\tilde{\cal E}$ is the set of unobserved lines within the island that are out. After inverting $\mathbf{B}$, this can be arranged in a matrix-vector form as follows,
	\begin{equation} \label{eq:factline}
	\begin{aligned}
	\tilde{\boldsymbol{\theta}} = \mathbf{B}^{-1} \mathbf{M s} +  \mathbf{B}^{-1} \tilde{\mathbf{p}}
	\end{aligned}
	\end{equation}
	where, $\mathbf{M}$ is the incidence matrix of the pre-disturbance island graph, see Section \ref{sec:model}. The vector $\mathbf{s}$ is sparse and its $i$th entry is $s_i:= \mathbf{m}^\mathsf{T}_i \mathbf{\theta}'/ x_i$, if $i$th line is disconnected, and $0$ otherwise. 
	
	\noindent\textit{Remark-I:} Following the initial outage, the pre-disturbance system's admittance matrix is invertible. However, approach in \cite{overbye}  is not directly applicable in the later stages of cascade as the overall system's in  admittance matrix is not invertible. The proposed approach avoids this issue by going island by island. 
	
	With the SCADA-based comunication network, the phase angles can not be measured. Therefore, in this paper, we convert the angle differences into line flow difference. Multiplying both sides of (\ref{eq:factline}) by $ \mathbf{D_{{\cal M}e} M^\mathsf{T}_{{\cal M}e}}$, we get,
	\begin{equation}
	\begin{aligned}
	\mathbf{D_{{\cal M}e} M^\mathsf{T}_{{\cal M}e}} \tilde{\boldsymbol{\theta}} = \mathbf{D_{{\cal M}e} M^\mathsf{T}_{{\cal M}e}} \mathbf{B}^{-1} \mathbf{M s} +  \mathbf{D_{{\cal M}e} M^\mathsf{T}_{{\cal M}e}} \mathbf{B}^{-1} \tilde{\mathbf{p}}
	\end{aligned}
	\end{equation}
		Here, subscript ${\cal M}e$ indicates quantities in measurable set and left hand side of the equality represents the change in power flow of the connected measurable lines $\tilde{\boldsymbol{\ell}}_{{\cal M}e} = [\boldsymbol{\ell}' - \boldsymbol{\ell} ]_{{\cal M}e}$. The problem of finding $\mathbf{s}$ can be formulated as the following optimization problem,
	\begin{equation} \label{eq:beforelassoeq}
	\begin{aligned}
	\hat{\mathbf{s}} : = arg \min_{\mathbf{s}}  \Big( ||\tilde{\boldsymbol{\ell}}_{{\cal M}e} - \mathbf{C} \tilde{\mathbf{p}} - \mathbf{C M s} ||^2_2 \Big)
	\end{aligned}
	\end{equation}
		Here, $C := \mathbf{D_{{\cal M}e} M^\mathsf{T}_{{\cal M}e}} \mathbf{B}^{-1}$. Different methods can be used to solve the aforementioned minimization problem. In this paper, a sparsity-based approach \cite{sparseidenti} is used. From (\ref{eq:beforelassoeq}), $\tilde{\boldsymbol{\ell}}_{{\cal M}e}  - \mathbf{C} \tilde{\mathbf{p}}$ is chosen as $\mathbf{y}$ and $\mathbf{C M}$ as matrix $\mathbf{A}$ to formulate the following LASSO,		\begin{equation} \label{eq:firstlasso}
	\begin{aligned}
	\hat{\mathbf{s}}^\lambda := arg \min_{\mathbf{s}} \Big(|| \mathbf{y - As} ||^2_2 + \lambda || \mathbf{s}||_1 \Big)
	\end{aligned}
	\end{equation}

	However, this does not incorporate any known breaker status information. For the lines, which we know are connected, the corresponding entries of $\mathbf{s}$ can be forced to $0$, (i.e., less than magnitude of element of $\mathbf{tol}$, which is used to detect the entries of $\mathbf{s}$ that are non-zero) using the matrix $\mathbf{A_{con}}$ as follows,
	\begin{equation} \label{eq:seclasso}
	\begin{aligned}
	\hat{\mathbf{s}}^\lambda := arg \min_{\mathbf{s}} \Big(|| \mathbf{y - As} ||^2_2 + \lambda || \mathbf{s}||_1 \Big)\\
	s.t. ~~ -\mathbf{tol} < \mathbf{A_{con} s} < \mathbf{tol} \hspace{30 pt}
	\end{aligned}
	\end{equation}
	
	\noindent\textit{Remark-II:} When estimating topology of the external area through PMU data, i.e., solving the problem in \cite{overbye}, equation (\ref{eq:factline}) can be used. In that situation,  $[\tilde{\boldsymbol{\theta}} - \mathbf{B}^{-1} \tilde{\mathbf{p}}]_{{\cal M}e}$ is chosen as $\mathbf{y}$, and $ [\mathbf{B}^{-1}]_{{\cal M}e}$ as matrix $\mathbf{A}$, for the LASSO in (\ref{eq:firstlasso}). 
	
		\vspace{-5pt}
	\subsection{Results}
	\begin{figure}
		\vspace{-25pt}
		\hspace{5pt}
		\includegraphics[scale = 0.58, trim= 1.2cm 0.2cm 1.3cm 0.2cm, clip=true]{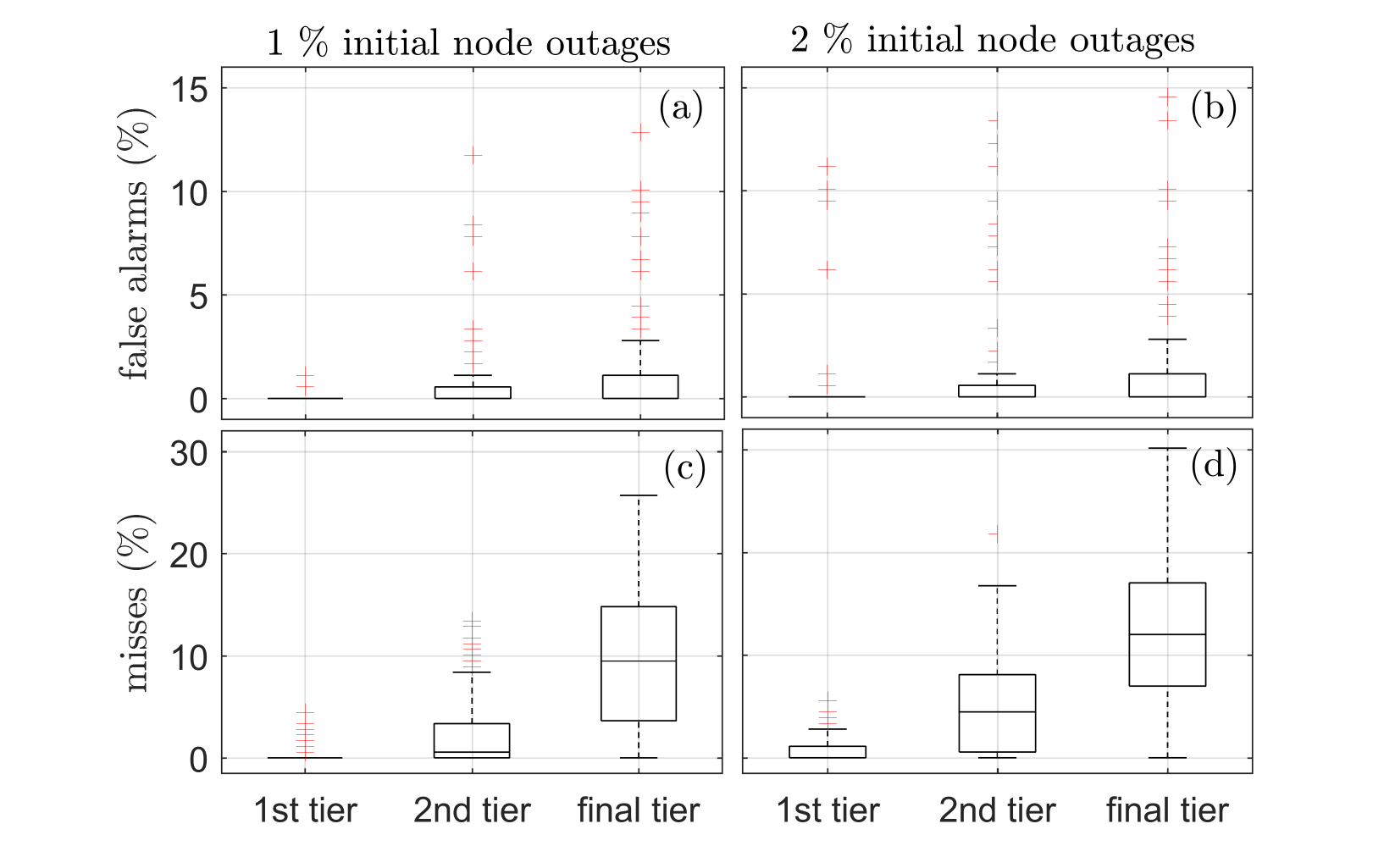}
		\vspace{-10pt}
		\caption{ Performance of line outage identification in IEEE 118-bus system: (a) and (b) Percentage of false alarms; (c) and (d) misses, following $1\%$ and $2\%$ of initial node outages, respectively. }
		\label{fig:falsepos}
		\vspace{-15pt}
	\end{figure}

\begin{figure}
		\vspace{-20pt}
		\hspace{5pt}
		\includegraphics[scale = 0.56, trim= 0.3cm 0.0cm 0.2cm 0.2cm, clip=true]{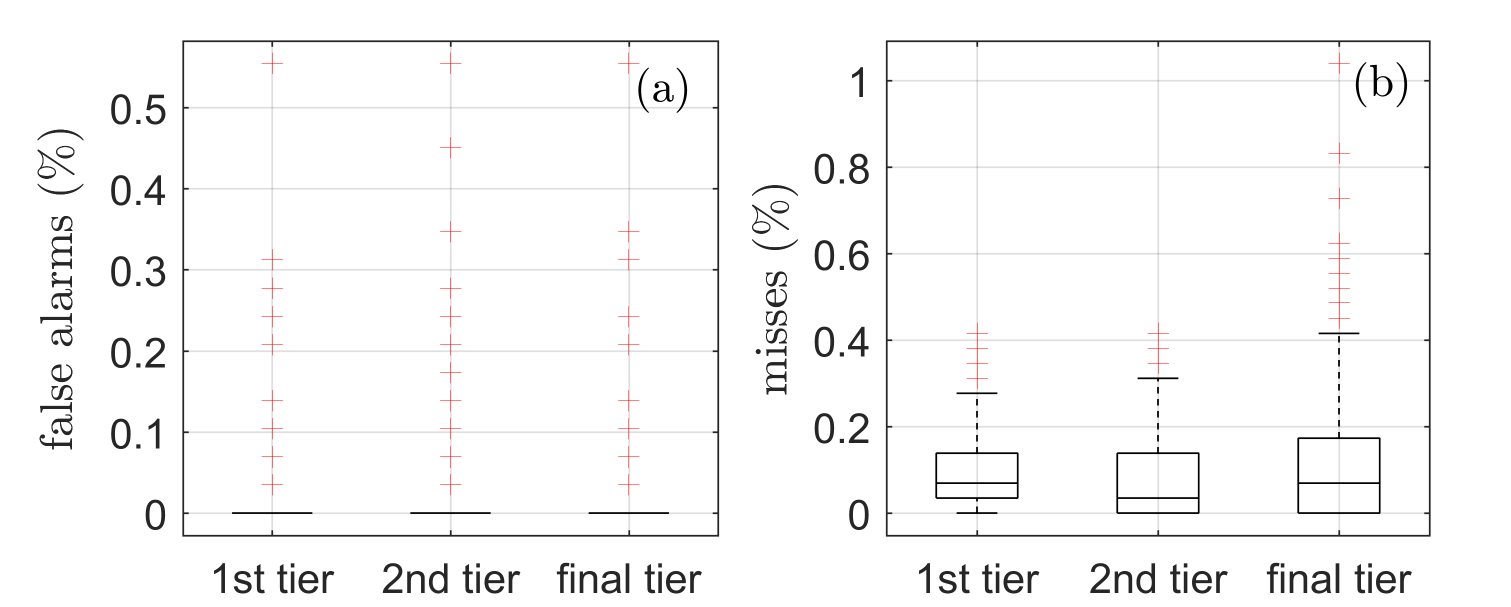}
		\vspace{-10pt}
		\caption{Performance of line outage identification, following initial $10$ node outages in Polish grid: Percentage of (a) false alarms and (b) misses.}
		\label{fig:falsepos_pol}
		\vspace{-20pt}
	\end{figure}

	In this section, we shall investigate how the effectiveness of bus and line outage estimations are impacted in the event of a cascading failure, wherein unlike Section \ref{sec:res:islandiden} the accuracy of the parent matters for further estimation. Like Section \ref{sec:res:islandiden}, the AC-QSS model is used for validation, and only real power variation is considered while measuring the change in line flow, to be consistent with the methodology. As the cascade progresses, the communication infrastructure is affected and the observability is degraded. Moreover, during the cascade an accurate identification of the parent becomes critical, since dependence on previous estimates causes an accumulation of error and further impacts the estimation. As shown in Fig. \ref{fig:falsepos}, the study performed in IEEE 118-bus system strengthens the understanding of the aforementioned error accumulation. Figures \ref{fig:falsepos} (c) and (d) illustrate the increase in percentage of misses, i.e., the cases where certain lines are out but are detected to be healthy, as the cascade progresses from one, two to the final tier of line outages. False alarms shown in 
	Figs \ref{fig:falsepos} (a) and (b) on the other hand do not seem to be affected as much by error accumulation and low observability.  This suggests that at a later stage of the cascade, the accumulation of error will cause our topology estimator to detect a subset of failed lines because the LASSO (\ref{eq:seclasso}) prefers sparse solutions. We also found that the results are hardly affected by noise. Figure \ref{fig:falsepos_pol} illustrates a similar pattern in the Polish system with initial node outage of size $10$. 
	For each percentage of initial nodal outage, $100$ and $500$ random cases were considered in the $118$-bus and the Polish system, respectively.

		\vspace{-3pt}
	\section{Preventive Control of Cascading Failure} \label{sec:Full_B}
	\vspace{-3pt}
	%
	
	As illustrated in Fig. \ref{fig:overall}, when the preventive controller located at CC is activated, it issues control commands that are communicated back to the actuators in power network. As a result, loss of communication leading to degrading controllability and observability will affect its performance.
	
		\vspace{-8pt}
	\subsection{Preventive Control Problem} The objective of the preventive controller is to stop cascade propagation through reduction of overloading in lines, which is defined as $\boldsymbol{\ell}_{ov}$. The objective is achieved by solving the following optimization problem \cite{prevent},
	\begin{subequations}\label{eq:Prev}
		\begin{align} 
	\hspace{-5pt} \min_{\mathbf{\Delta \theta},\boldsymbol{\ell}_{ov}, \mathbf{\Delta p}_{g,l}}  & -\mathbf{1}^\mathsf{T} \mathbf{\Delta p}_l+ \mathbf{\lambda}^\mathsf{T} \boldsymbol{\ell}_{ov}  \\
	\textrm{s.t.} ~~~ & \mathbf{\Delta p}_g - \mathbf{\Delta p}_l = \mathbf{\hat{B}\Delta \theta}, \Delta \theta_i = 0, \forall i \in \Omega_{ref} \\
	 |\boldsymbol{\ell}_{{\cal M}e} + &\mathbf{D_{{\cal M}e} M^\mathsf{T}_{{\cal M}e}}  \mathbf{\Delta \theta} | \leq \boldsymbol{\ell}_{max} + \boldsymbol{\ell}_{ov}, ~ \boldsymbol{\ell}_{ov} \geq \mathbf{0}  \label{Prev:absolute}\\
	& -(\mathbf{p}_g)_{{\cal M}e}  \leq (\mathbf{\Delta p}_g)_{{\cal M}e}  \leq \mathbf{0}   \\
	& -(\mathbf{p}_l)_{{\cal M}e}  \leq (\mathbf{\Delta p}_l)_{{\cal M}e}  \leq \mathbf{0} \\
	& (\mathbf{\Delta p}_g)_{ \overline{{\cal M}e}} = \mathbf{0}, (\mathbf{\Delta p}_l)_{ \overline{{\cal M}e}} = \mathbf{0}
	\end{align}
	\end{subequations}
	Here, $\Delta$ represents  changes in quantities, and the elements of vectors $\mathbf{p}_g$ and $\mathbf{p}_l$ represent the generation and load at each bus, respectively. 
The matrix $\mathbf{\hat{B}}$ is the estimated admittance matrix of the network. The variables $\boldsymbol{\ell}$ and its allowable $\boldsymbol{\ell}_{max}$ represent the actual power flow and its maximum value, respectively. 
Subscript ${\overline{\cal M}e}$ represent quantities in the unobservable set.
	As presented in Fig. \ref{fig:overall}, $\mathbf{\hat{B}}$, $\boldsymbol{\ell}_{{\cal M}e}$, $(\mathbf{p}_g)_{{\cal M}e}$, and $(\mathbf{p}_l)_{{\cal M}e}$ are taken as inputs to the optimization problem, and the solution gives load shedding set points and generation re-dispatch commands as outputs. 
	
	\vspace{-7pt}
	\subsection{Results}
		\vspace{-5pt}
		
	\begin{figure}
		\vspace{-20pt}
		\hspace{-8pt}
		\includegraphics[scale = 0.45, trim= 0.7cm 0.2cm 0.6cm 0.1cm, clip=true]{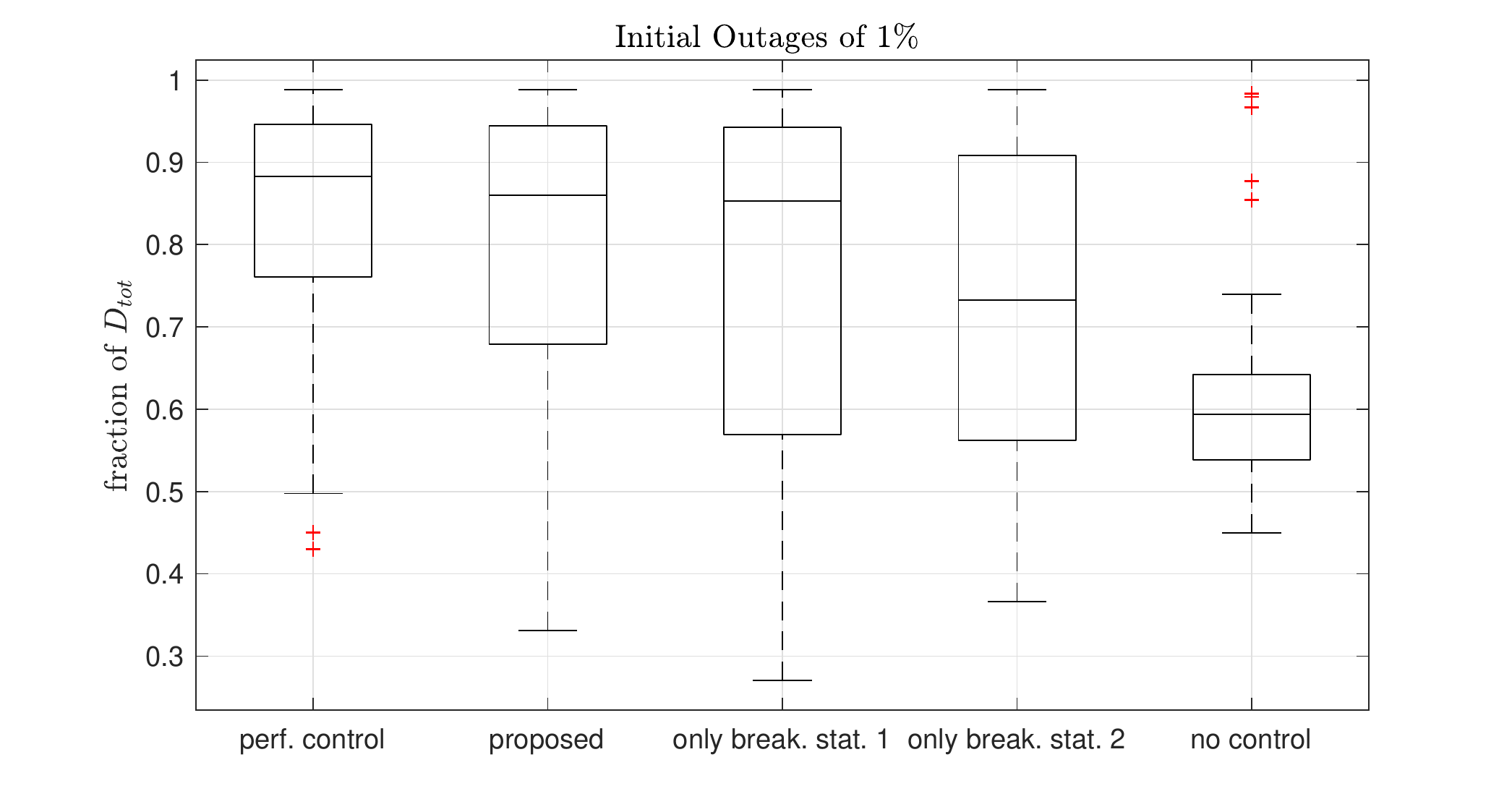}
		\vspace{-25pt}
		\caption{Boxplots of total load served at the end of cascade following $100$ random initial failures of $1\%$ of nodes in IEEE $118$-bus system. Comparison between preventive control with proposed method ($\mathbf{\hat{B}}$ from proposed estimation), perfect control ($\mathbf{B}$ known), control using $\mathbf{B}$ constructed from partial breaker statuses, and no preventive control.}
		\label{fig:1per118}
		\vspace{-15pt}
	\end{figure}
	
	The preventive controller in \cite{prevent} was validated using a DC-QSS model. The same model is used in this section, to evaluate the effectiveness of proposed estimation in the closed-loop system. For the sake of comparison the following cases are considered, \\
		\underline{\emph{(i) Perfect control}}: The topology processor and preventive controller have complete observability and controllability, respectively, of the power network assuming a perfectly resilient cyber layer. We expect this case to have the best performance. \\
		\underline{\emph{(ii) Proposed}}: The cyber layer is impacted by 
		coupled cascading failure resulting in progressively reduced observability and controllability, and the $\mathbf{\hat{B}}$ matrix obtained by proposed estimation process is used in the controller.\\
		\underline{\emph{(iii) \& (iv) Only breaker status 1 \& 2}}: This is same as case (ii) with proposed estimation process turned off. For case (iii), we assume all the unobservable lines are operational while constructing the $\mathbf{{B}}$ matrix. In contrast, for case (iv) we assume all the unobservable lines are not operational.\\
		\underline{\emph{(v) No control}}: The preventive controller is turned off and the cascade progresses naturally without any influence from the cyber layer. Ideally, this case needs to be outperformed by all.
	
	Figure \ref{fig:1per118} shows the boxplots of total load served at the end of cascade in each of these cases following $100$ random initial failures of $1\%$ of nodes in IEEE $118$-bus system. The proposed method performs better than cases (iii), (iv) and (v), and falls just short of case (i) w.r.t. the median, the $25$th and $75$th percentiles. Only for a few extreme cases this performs worse than having no control. Comparatively, case (iii) and (iv) with wrong $\mathbf{B}$ matrix performs worse than having no control at all. This emphasizes the need for deployment of better estimation tools when implementing control to contain the cascade.

	\section{Conclusion} \label{sec:con}	
	A new approach was proposed in this paper that no longer requires the assumption of connectedness in estimating line outages. In this method, first an integer linear programming problem was formulated to segregate the buses into various connected components. Then, changes in boundary line power injections and redispatch of generation and load were incorporated into the classical line outage identification problem. Further, the observable breaker statuses were integrated as constraints into this 
	problem. The effectiveness of estimation process was studied as coupled cascading failure propagated in power and communication systems. 
	It is shown that due to shrinkage of the observable areas and the reliance on previous estimates, error accumulated and negatively impacted the estimation accuracy. Finally, the estimated admittance matrix was fed into the preventive control algorithm of cascading failure. Central tendency and dispersion measures of per unit total demand served at the end of cascade show a superior performance of the proposed method compared to cases where it is turned off.
	\vspace{-0pt}
	%
	\bibliographystyle{IEEEtran}

\end{document}